\documentstyle[12pt]{article}
\newcommand{\eg}{{\sl e.g.}}
\newcommand{\ie}{{\sl i.e.}}

\newcommand{\etal}{{\sl et al.}}

\newcommand{\ltae}{\raisebox{-0.6ex}{$\,\stackrel
{\raisebox{-.2ex}{$\textstyle <$}}{\sim}\,$}}
\newcommand{\gtae}{\raisebox{-0.6ex}{$\,\stackrel
{\raisebox{-.2ex}{$\textstyle >$}}{\sim}\,$}}

\begin{document}
\large
{\bf Nuclear bars and blue nuclei within barred spiral galaxies}

\vspace{1.5cm}
\normalsize
{\bf M. Shaw$^*$}

\small
{\it Dept. of Physics, University of Sheffield, The Hicks Building,
Sheffield, S3 7RH.}

\vspace{0.7cm}
\normalsize
{\bf D. Axon$^{* \ddag}$}

\small
{\it Space Telescope Science Institute, 3700 San Martin Drive,
Baltimore, MD21218, USA.}

\vspace{0.7cm}
\normalsize
{\bf R. Probst \&\ I. Gatley}

\small
{\it Kitt Peak National Observatory, P.O. Box 26732, Tucson, AZ85726, USA.}

\normalsize
\vspace{1.0cm}
Running title : Nuclear bars and blue nuclei within barred spiral galaxies.

\vspace{1.0cm}
Address to send proofs : Dept. of Physics, University of Sheffield.

\vspace{1.0cm}
Send offprint requests to : Martin Shaw

\vspace{1.0cm}
Thesaurus codes : 11.01.2, 11.14.1, 11.16.1, 13.09.1

\vspace{1.0cm}
Accepted for publication in Mon. Not. R. astr. Soc.

\vspace{3.0cm}
* Visiting astronomer, Kitt Peak National Observatory, NOAO, which is
operated by AURA, Inc. under contract to the US National Science Foundation.

\ddag On leave from Nuffield Radio Astronomy Laboratory, University of
Manchester, Jodrell Bank, Macclesfield, Cheshire, SK11 9DL.

\normalsize
\newpage
\baselineskip 20pt

\parindent 0.0cm

{\bf Summary.}
\parindent 1.0cm

Multicolour near IR photometry for a sample of 32 large barred spiral galaxies
is presented. By applying ellipse fitting techniques, we identify significant
isophote twists with respect to the primary bar axis in the nuclear regions of
$\sim$70 \%\ of the sample. These twists are identified in galaxies as late as
SBbc and are clearly distinguishable from spiral arm morphology. At most seven
of the
galaxies with isophote twists are inferred to possess secondary
(nuclear) bars, the axis ratios of which appear to correlate with morphological
type. The remainder may result from triaxial bulges, or from oblate bulges
misaligned with the primary bar.

The near IR colour distributions in these data show evidence for (red)
circumnuclear star forming rings in 4 galaxies. The majority of the sample
(19) also possess striking blue nuclear regions, bluer than typical
old stellar populations by $\sim$0.3 mag. in (J--H) and $\sim$0.23 mag. in
(H--K). Such blue colours do not appear to correlate with the presence of
nuclear rings or pseudo--rings, nor with the activity of the host galaxy (as
determined from emission--line spectroscopic characteristics). Several
mechanisms to explain this blue colour are considered.

\vspace*{2.0cm}
\begin{description}
\item [Keywords:] galaxies: active -- galaxies: nuclei -- galaxies: photometry
-- infrared: galaxies
\end{description}

\newpage
\section{Introduction}

The presence of a bar is likely to be an important factor in the fuelling
of nuclear (starburst/AGN) activity in many galaxies (\eg\ Heckman 1980,
Hawarden \etal\ 1986, Arsenault 1989). An increasing proportion
of barred galaxies show evidence of starbursts in the form of active star
formation within circumnuclear rings -- CNR's. Good examples of this
phenomenon are NGC1097, 4321 and 5728 (Shaw \etal\ 1993 -- henceforth Paper I).
This is readily understood
as a direct result of gas motions within the barred potential:
the bar provides an efficient mechanism to
transport gas from the spiral arms to the inner Lindblad resonance (ILR) --
or the outer ILR if two exist -- where the resulting reservoir of gas provides
a
source for star formation activity.

Recent theoretical studies (\eg\ Shlosman \etal\ 1989, Pfenniger \&\ Norman
1990) have also highlighted a mechanism whereby an independent,
rapidly rotating nuclear bar can provide an efficient means of transporting
material from the CNR onto the nucleus, thereby possibly fuelling LINER/Seyfert
activity. Some support for this hypothesis may exist in those barred
Seyfert galaxies
which also possess CNR's (\eg\ NGC1097: Hummel \etal\ 1987b; NGC3783: Winge
\etal\ 1992;
NGC5728: Schommer \etal\ 1988, Wilson \etal\ 1993; NGC6951: Boer
\&\ Schultz 1993). Shaw \etal\ (Paper I) stress the
significant r\^{o}le any such mass transfer of gas may have on the
stellar dynamics in the nuclei of barred galaxies.

Inevitably, any study of the fuelling of nuclear activity by a bar
requires detailed knowledge of the mass distribution in the central
region. In general
such information is unavailable as existing observational constraints on
the stellar dynamics in the nuclear regions of barred potentials are
lacking. The influence of obscuration by dust in the nuclei of barred
galaxies (\eg\ Baumgart \&\ Peterson 1986) is also a problem, as this places
a fundamental limit on photometric studies of the inner regions. The
problem is particularly acute given the large number of barred galaxies
anticipated to possess nuclear isophote twists suggestive of nuclear bars
or coupled star/gas components (Paper I).  An important benefit of using
near IR photometry is that the effects of dust obscuration are greatly
reduced, leading to a more reliable determination of the morphology
of the underlying old stellar populations, and ultimately the gravitational
potential, in the nuclear regions. No near IR survey of the nuclear regions
of barred spiral galaxies has been undertaken.
It is thus unclear what proportion of barred galaxies possess nuclear bars,
and whether there is indeed the strong correlation between such bars and the
presence of CNR's implied in Paper I. The degree to which nuclear bars can
be distinguished from possible triaxial bulges
(\eg\ Kormendy 1979, 1982) is also unknown.

In this paper, we present the results of a near IR photometric study of
32 nearby barred spirals which seeks to address such issues. For the reasons
given above, we are able to extend studies of the nuclear properties of
barred galaxies to significantly later morphological types than has been
possible to date.
The specific objectives in
our study are to determine the proportion of barred galaxies which possess
nuclear bars, CNR's and/or isophote twists in a statistically significant
sample. In searching for CNR's, we make use of the sensitivity of the K--band
(2.2 $\mu$m) to emission from M giants/supergiants residing in regions of
ongoing star formation activity. Analyses of near IR colour distributions are,
therefore, an important aspect of this study. Given the limitations inherent
in the use of rotation curves to define the pattern speed of a bar, the
identification and properties of CNR's take on added importance if, as is
suggested by N--body simulations (\eg\ Combes \&\ Gerin 1985),
they identify the location of one
of the principal resonances in a barred potential.

\section{The sample, data acquisition and reduction}

\subsection{Sample selection}

The data in this paper comprise near IR images of 32 large spiral
galaxies classified as SAB or SB by de Vaucouleurs \etal\ (1976). Table 1
collates general properties of these objects.

The present dataset constitutes a subset of our complete sample of large
(D$_{25}$ \gtae\ 2 arcmin), bright (m$_B$ \ltae\ 12 mag.), non--interacting nor
irregular barred galaxies. This sample was constructed primarily from Hummel
\etal\ (1987b), Arsenault (1989 -- both control and AGN/STB samples),
Kormendy (1979), Pogge (1989), Buta \&\ Crocker (1993) and Friedli (private
communication). It therefore includes all nearby
barred galaxies which possess nuclear rings,
pseudo--rings, lenses, that show excessive nuclear H$\alpha$ fluxes or
evidence for nuclear activity (starburst, LINER or Seyfert). The size criterion
is chosen so as to improve spatial resolution in the nuclear regions.

The sample presented in this paper
is not complete, as it constitutes the first observations acquired from our
sample. More general discussions, \eg\ relating to the statistical properties
of the sample, will be presented in future papers as and when the observational
data become available. Future data will improve the available spatial
resolution by utilising alternative pixel scales. However,
the number of objects studied in the present paper is sufficient to
constitute a statistically significant subset and thus allows
an exploratory investigation of the frequency of nuclear bars, isophote
twists and
investigating the possible links with CNR's. The subsample does contain an
inherent bias insofar as there was a preferential selection of the
larger (though not necessarily more face--on) barred galaxies. However,
we stress that an important aspect
of this subsample is the broad range of morphological types represented.

\subsection{Data acquisition and reduction}

Simultaneous J, H, and K (1.2 -- 2.2 $\mu$m) photometry of the present
sample was undertaken using the (256x256 element PtSi array) multichannel
IR camera (Ellis \etal\ 1992) on
the KPNO 1.3m between 5 and 14 March 1992. The pixel scale in this
configuration is 1.35 arcsec, or 0.10--0.45 kpc/pixel over the distance
range of our sample. Integration times of 3x180 sec
on each object were interspersed by equal length sky frames, typically offset
from the galaxy by $\sim$10 arcmin. Each object frame was also offset
from the other by $\sim$10 arcsec to minimise contamination by defective
pixels. Dark current subtraction was facilitated using equal length dark
frames through the run.

Any contaminating objects were removed from the sky offset images, and the
resulting frames median filtered to create high S/N ``flat--field'' frames
for each galaxy in each colour. Such images were normalised to a mean of
unity and were divided into the corresponding object frame for that
particular filter to remove pixel--to--pixel response variations. Such
flat--fielded frames were then spatially aligned
and summed. Residual edge--effects were removed
by subsetting the images and any remaining defective pixels removed by
interpolation.

The object frames were then calibrated using observations of 10 faint (J
= 6.6--8.4 mag) near IR standards derived from Elias \etal\ (1982).
Typically, between 4 and 8
stars were observed each night. Four of the five nights proved to be highly
photometric -- the derived zeropoints being consistent throughout
the night to within $\pm$0.01--0.02 mag. arcsec$^{-2}$ in all passbands.
Data from the single non--photometric night, which affects observations of
NGC4754 alone, were calibrated using observations of the standard star HD129653
observed immediately prior to the object frames. Differential airmass
corrections were applied to the derived zeropoints using mean extinction
coefficients appropriate for the site.
The sky background estimates used in each case were median values derived
from the object frames directly (in the majority of cases where sky
dominates the frames), or from the median filtered sky offset frames (where
the objects fill the frame). In the latter, as no significant sky background
variations were noted between each sky frame acquired, the mean flux
from each of the three frames provides a good estimate of the sky for the
objects
concerned.
Since the J, H
and K images were acquired simultaneously, no correction for relative
atmospheric extinction differences was necessary when deriving the colour
distributions of each object.

The seeing, measured directly from the object frames throughout the run,
was 2.0 ($\pm$0.3) arcsec FWHM at J,
1.9 ($\pm$0.1) arcsec at H and 1.9 ($\pm$0.2) arcsec at K. In consequence,
ellipse--fitting results derived from the innermost $\sim$2 arcsec must be
considered highly unreliable.

\section{Results}

\subsection{Primary bar position angles}

An important first step in searching for evidence of isophote twists in the
nuclei of barred galaxies is an assessment of the orientation of the primary
bar.

In each object, visual estimates derived from
contour plots and greyscale images of each J, H, K image were compared to
those derived from the ellipse fitting
procedures discussed below. This is necessary because the latter technique
can be unduely influenced by spiral arm morphology (section 3.2). The
respective
measurements are listed in Table 2, together with their corresponding
uncertainties. [Absolute N and E were defined from the known N--S
and E--W positional offsets of the standard star observations.]
The quoted errors reflect the uncertainty in a particular
position angle estimate, and the spread in position angles derived from
each passband. Also listed in
this table are angles measured from blue and/or red sky survey plates in those
objects for which the principal bars can be clearly defined.

Of the 32 objects listed, 3 are too edge--on to allow determination of the
primary bar position angle. Such objects will not be discussed further in the
context of identifying possible isophote twists.
Measures for the primary
bar position angles for the remainder
are clearly equivalent between the present data and sky survey plate
estimates in all cases other than NGC3953
and NGC4536. Both galaxies are highly inclined systems (i $>$ 60$^{\circ}$,
90$^{\circ}$ corresponding to edge--on), compromising the evaluation
of the position angle from the sky survey plates. In these cases, the measures
derived from the present data have been adopted.

\subsection{Ellipse fitting}

Initial inspection of the data revealed clear evidence for isophote twists
(with respect to the primary bar position angles) in 16 objects. With a view
to quantifying this, and measuring the
degree of isophote misalignment more explicitly, ellipse fitting techniques
(\eg\ Jedrzejewski 1987)
were applied to each (J, H, K) image independently. As a consequence, 21
objects (72\%\ of the non edge--on sample) were found to possess measureable
isophote twists, the magnitudes of which are collated in Table 2. [This
frequency does not include NGC4321 which possesses elongated
inner isophotes and will be discussed further below.] Such a
high incidence of isophote misalignment confirms the expectations in
Paper I: this is a  common property of the central near IR light
distributions of barred galaxies of all
morphological types. Figure 1 illustrates this with the 4 most
striking examples of isophote twists (NGC3941, 3953, 4613 \&\
4754), whilst the results of the ellipse fitting to these particular
objects are shown in Figure 2.

On the assumption that spiral arms are trailing, we have made an assessment
of whether the measured isophote twists trail or lead the principal bar in
a similar fashion to that undertaken by Buta \&\ Crocker (1993). Our
assessment is illustrated by the sign of the twists in Table 2. The magnitude
of the twists, and spatial regions over which they occur, are broadly
consistent between each passband (the differences being reflected in the
associated errors in this table). Only in the outer regions of some objects
are discordant ellipse fitting results derived. In part, these are due to
intrinsic colour variations within the disc components (manifest in the case
of NGC4321 -- Figure 4), but are also a reflection of the impact of the
declining quantum efficiency variations of the IR array with wavelength
(falling from 6.6\%\ at J to 3.4\%\ at K). The limited spatial coverage of the
disc component may be an additional factor for the larger galaxies.

Undoubtedly,
NGC4321 and NGC4274 present the clearest evidence of a nuclear bar within the
present
sample as Figures 3 and 4 clearly show. The nuclear bar in NGC4321 extends to a
radius $\sim$9 arcsec (1.3 kpc) -- witness the trough in ellipticity, and peak
in $\cos 4\theta$, profiles at this point (Figure 4).
This bar possesses an axis
ratio similar to that of the main galaxy at large radius and is aligned with
the
inferred primary bar major axis to within 7$^{\circ}$
(adopting a position angle from Arsenault
\etal\ 1988 and references therein). It is also immediately interior
to the CNR evident in
our data (Section 3.3) and in emission--line images (Arsenault \etal\ 1988).
Thus, in some respects the properties of the central $\sim$1.5 kpc of
NGC4321 bear a striking similarity to equivalent regions in NGC1097 and
NGC5728 (Paper I). Unlike those objects, however, the inner bars in NGC4321
and NGC 4274 are closely aligned with the primary bar.

A correlation exists between the scale of the measured
isophote twists ($\delta$R$_{max}$) and bar length -- larger bars
displaying isophote twists over larger radii (the derived product--moment
correlation
coefficient of 0.68 is significant at $>$ 99\% significance level). An
equally significant correlation between $\delta$R$_{max}$ and morphological
type results from
the fact that bars in the later type galaxies in this sample tend to be
larger. [Note that the anti--correlation between bar and bulge dimensions
found by Athanassoula \&\ Martinet (1980) was restricted to objects earlier
than Sa.]

Interestingly,
the nuclear bar axis ratios in NGC1097, NGC4274, NGC4321 and NGC5728
are markedly smaller than
those identified in early type galaxies (\eg\ Jarvis \etal\ 1988),
suggesting a possible dependence on
morphological type or, equivalently, the nature of the primary bar.

\subsection{Colour distributions}

A striking feature of the derived colour maps is that 19 of the 32 galaxies
show particularly blue nuclei, \ie\ regions of the colour maps which
are considerably bluer than those areas
of each galaxy dominated by the bulge component. We refer to these objects
as comprising the ``blue nuclei'' sample. Examples of these structures
are given in Figure 5 (b--e). We quantify the
magnitude of these colours in Table 3. [The errors quoted throughout
this section correspond to standard errors on the mean values for a particular
dataset.]

As is evident in this table,
a continuum of colours exist between these ``blue nuclei'' sample
and the remainder
(henceforth referred to as the ``control'' sample). The
distinguishing property of the former sample is the
identification of substructure within the colour maps.
Typical nuclear colours within a 0.38 ($\pm$0.02) kpc radius aperture are
0.45 ($\pm$0.06) mag. in (J--H) and 0.02 ($\pm$0.04) mag. in (H--K) for the
``blue nuclei'' sample.
The colours of the ``control'' galaxies
within equivalent apertures are 0.59 ($\pm$0.06) mag.
and 0.19 ($\pm$0.04) mag. respectively. Interestingly, even the ``control''
sample
colours are somewhat bluer than those of typical spiral galaxies:
the median colours (\ie\ averaged
over all morphological types) from Griersmith \etal\ (1982) are 0.71--0.76
mag. in (J--H) and 0.25--0.30 mag. in (H--K). Similarly, within 6 arcsec
diameter apertures, data from Forbes \etal\ (1992) yield typical colours of
(J--H) = 0.83 ($\pm$0.14) mag. and (H--K) = 0.44 ($\pm$0.16) mag. for a sample
of 15 galaxies, only two of which are barred. An IR colour--colour diagram
for the galaxies listed in Table 3 is shown in Figure 6, where
each galaxy has been identified according to its spectroscopic (nuclear
activity)
characteristics as presented in Table 1.

The colour maps (\eg\ Figure 5) also identify the presence
of the CNR's in
NGC4303, 4314 and 4321. These rings are generally redder than all other regions
in the galaxy, having colours consistent with those of typical old stellar
populations in ellipticals and spiral bulges (Table 4). However, given the
spatial distribution of the near IR emission in the rings, a significant
contribution from M giants/supergiants in regions of active star formation
is likely. This is consistent with the situation in NGC1097 and 5728, where
an excellent spatial correspondence is seen between the 2.2$\mu$m luminosity
peaks and H$\alpha$ emission within each CNR (Paper I).
The inferred radii of the rings identified in the present data (Table 4) are in
good agreement with those measures tabulated by Pogge (1989) and Arsenault
\etal\ (1988). The ring colours we derive for NGC4314 agree with those
measured by Benedict \etal\ (1992).

Despite its inclusion in a list of galaxies possessing H$\alpha$ nuclear rings
(Pogge 1989), NGC3351 only shows evidence for a CNR within our (J--K)
colour map --
Figure 5 (a). The brightest emission regions within this ring have (H--K)
$\sim$0.37 mag. and (J--K) $\sim$0.99 mag., the latter value being equivalent
to that of the nucleus itself. Moreover, the major axis of this ring
is aligned with the twisted isophotes identified in Table 2. Both the ring
orientation and the location of the maxima in (J--K) coincide exactly with the
structure seen in CO and discussed by Kenney \etal\ (1992).

We have investigated a possible link between CNR's and nuclear colours
by comparing the objects in Table 3 with the catalogue of nuclear
rings and pseudo--rings in Buta \&\ Crocker (1993). Although,
three of the galaxies
in our ``blue nuclei''
sample also have nuclear rings/pseudo--rings,
two of the galaxies in the control sample also show evidence of such
rings, implying little evidence of a correlation. Of course, such statistics
are only suggestive given the (distance dependent) selection effects inherent
in our data and those of Buta \&\ Crocker. Indeed, one must also consider
the possibility that the nuclear colours in the majority of galaxies in Table 3
are contaminated by
CNR's which remain unresolved in our data. This is particularly
important given that ILR's -- supposedly identifying the presence of CNR's
(Combes \&\ Gerin 1985) --
show a wide variation in size within the N--body simulations
conducted to date (\eg\ a factor of two in Paper I). There also exists the
possibility of temporal variations in the size of CNR's (Combes \etal\ 1992).

Rings in
our data would remain unresolved if their radii were $\sim$ 3 pixels or less,
corresponding to $\sim$0.5 kpc since the distance to those blue nuclei
galaxies in Table 3 not identified as possessing CNR's is typically
28.0 ($\pm$1.5) Mpc. The possibility therefore remains that the majority
of colours in Table 3 are influenced by the presence of unresolved CNR's.
Clearly,
since the CNR's unambiguously identified in the present dataset are very red
(Table 4), their presence in the remaining data would imply that even bluer
underlying
colours exist within these nuclear regions.
Possible sources of such blue colours are
discussed in Section 4.2 below.

\section{Discussion}

\subsection{Nuclear bars and triaxial bulges}

A primary aim of the present study was to quantify the occurence of
isophote twists in the nuclear regions of barred galaxies. As Table 2
graphically illustrates, such twists are common in the near IR. Fully
$\sim$70 \%\ of the non edge--on galaxies in the present sample display
measurable twists in galaxies ranging from SB0 to SBbc.
It is therefore important to identify the mechanism(s)
by which such twists arise. Since contamination by spiral arms is minimal
in the regions concerned, nuclear bars or misaligned/triaxial bulges present
the most obvious cause.

The nature and properties of nuclear bars have been investigated in several
theoretical studies (\eg\ Shlosman \etal\ 1989, Pfenniger \&\ Norman 1990).
Only two
investigations have used constraints imposed by observations of real galaxies.
In Paper I, Shaw \etal\ argued that twists result from
the dynamical influence of the gas component (on the stars) as the gas follows
the stable x2 orbits immediately interior to the ILR. Conversely, Friedli \&\
Martinet (1993) view these systems as nested double barred galaxies. A
principal
difference between the simulations conducted in these studies is the r\^{o}le
of
dissipation. In Paper I viscosity is the mechanism by which the gas leads the
stars -- giving rise to the observed twists. Dissipation is
less important in the picture described by Friedli \&\ Martinet (1993). In
their scheme,
the twists are a direct manifestation of two misaligned bars possessing grossly
different pattern speeds.

We have identified what proportion of those galaxies displaying
isophote twists (other than NGC4321)
are candidates for possessing nuclear/misaligned bars. In doing so, three
methods are available to us. In the first instance,
we have compared the spatial dimensions of the observed twists
($\delta$R) to the
lengths (L) of the principal bars (Table 2). In the mean, $\delta$R/L
extends from 0.16 ($\pm$0.02) to 0.67 ($\pm$0.05), neglecting NGC4321 -- \ie\
considerably in excess of the 0.1--0.2 inferred by Friedli \&\ Martinet (1993)
or the values of 0.14 and 0.09 observed in NGC1097 and 5728 respectively
(Paper I). The only likely nuclear bar candidates on these grounds would be
NGC4643 and, possibly, NGC4274.
Moreover, in the hypothesis advanced in Paper I, the observed twists
must lie immediately interior to the radius of the CNR. Again neglecting
NGC4321, the mean dimensions of the observed twists have inner and outer
radii of 0.8 ($\pm$0.1)
kpc and 2.9 ($\pm$0.3) kpc respectively, seemingly larger than the typical
radii of CNR's (Paper I and references therein).

A second discriminant comes from the misalignment angles in Table 2. In
agreement
with the (limited) sample presented by Buta \&\ Crocker (1993), the isophote
twists lead or trail the spiral arms in roughly equal proportions, even for
the most face--on galaxies in the sample. [The measures in Table
2 have not been deprojected as the measured  inclination angles
are derived from projected axis ratios and are notoriously unreliable --
depending, for example, on spiral arm morphology.] This suggests that
relatively
few of the objects in Table 2 are candidates for double barred galaxies,
although a more reliable conclusion would require accurate deprojection.

Finally, we consider the orbital families in the barred potential. Stable
periodic orbits cannot cross (Sparke \&\ Sellwood 1987, Athanassoula
1992). Therefore, the minor axis dimension of the primary bar (identifying the
x1 orbital family) corresponds to the maximum dimension of any nuclear bar and
thus the radius of any CNR. We have determined the axis ratios of the
principal bars from the J--band images of each galaxy with measurable isophote
twists in Table 2. No deprojection was undertaken -- we assume the nuclear
and primary bars are coplanar and that projection foreshortens the dimensions
of each equally. Comparison of the minor axis dimensions to the radial scales
of the twists listed in this table suggests that nuclear bars are present in
NGC4262, 4274, 4314 and 4643, with NGC4371 and 4754 being additional, though
less
well--defined, candidates.

In conclusion, at most 7 of the 22 objects displaying near IR isophote twists
are likely to possess nuclear/misaligned bars of the type envisaged in
Paper I or Friedli \&\ Martinet (1993). Using the present data, a
discrimination
between these two studies would require a detailed investigation of the
r\^{o}le
of projection. As one must take into account the intrinsic figure of the
bar itself (Paper I), this analysis is beyond the scope of the present paper.
Observationally, a more definitive conclusion would result from detailed 2D
stellar and gas dynamics of galaxies possessing twisted isophotes.

It is possible that the observed twists in the remaining galaxies
result from the presence of triaxial bulges (\eg\ Kormendy 1979, 1982).
However, the
complex transition between bulge and primary bar luminosity components renders
unreliable any results derived from photometry alone. In fact, the observed
twists could readily result from a misalignment between a conventional (oblate)
bulge and the principal bar. The stability of such a configuration remains to
be determined, although the frequency of twists in the present sample implies
such a configuration must be stable. A definitive identification of triaxiality
in these galaxies could only be undertaken by combining full 2D luminosity
decomposition with extensive stellar and gas kinematics.

\subsection{The significance of blue nuclear colours}

Many galaxies in the present sample display blue nuclear colours. Indeed,
even in those galaxies displaying CNR's,
blue nuclei are evident on scales considerably smaller than the inferred
dimensions of rings (Figure 5).
We consider two mechanisms by which this blue colour
could arise.

\subsubsection{Extinction variations}

The first results from a {\it reduction}
in extinction within the
nuclear regions. The reddening vector in Figure 6 suggests a change of
A$_v$ $\sim$1 mag. would be sufficient to bring the IR colours of the majority
of the sample into agreement with those of typical old stellar populations.
Reduced extinction may result from the highly collimated nature of the gas
distribution in a bar. The gas is principally restricted to the leading edges
of the bar (as in, for example, NGC613: Hummel \etal\ 1987a, and NGC1097:
Gerin \etal\ 1988). The region interior to the corotation radius, but outside
the bar itself, is thus largely devoid of gas in many barred galaxies.
An aperture $\sim$0.4 kpc in
radius (Section 3.3) would encompass such regions, possibly leading to a
reduction in the near IR colour indices when compared to non--barred galaxies.
Optical colours would not support this assertion, since the nuclear colours
derived are invariably very red. However, large scale surveys of near IR
galaxian colour distributions remain to be undertaken.

\subsubsection{Nuclear activity}

The blue colour could be an observational consequence of nuclear
activity if it implies the existence of starbursts with suitable star formation
properties. This hypothesis is favoured by the fact that, of the 13 galaxies in
the sample whose nuclei have the spectral characteristics of starbursts/LINERS
(Arsenault 1989 and references therein), 10 have unusually blue nuclei.
Conversely, only 9 galaxies not showing such activity also possess blue
colours. Formally, sampling theory suggests that these ratios differ at the
93.6\%\ significance level using the ``two--tailed'' test. In
reality, however, the results are only suggestive given the limited
samples involved and the large apertures used in the spectral
classification (4--8 arcsec diameter -- Keel (1983) -- such that much of the
emission from the starburst candidates may come from CNR's rather than the
nuclei directly).

A link between blue colours and nuclear activity could be understood as
resulting
from the featureless blue continuum (FBC) detected in many Seyferts
and LINERS (\eg\ Yee 1983). This FBC is bluer (at optical wavelengths)
than the colours of spiral galaxy bulges, and may correspond to the
continuum in a young starburst (Terlevich \&\ Melnick 1985). A difficulty with
this
hypothesis results from the fact that
the blue colours in Table 3 are often distributed in a collimated structure
(Figure 5), and thus are more extended than the (unresolved) FBC in most
active galaxies. Moreover, the contribution of the FBC is likely to be small
in the near IR. The blue colours in our sample are observed even in galaxies
which are not classified as active whilst, in the scheme of Terlevich \&\
Melnick (1985), ``blue LINERS'' are galaxies which have already evolved beyond
a Seyfert 2 phase of activity.

There also exists the intriguing
possibility of a relationship between the blue nuclear
structures (Figure 5) and the anisotropic radiation field in active galaxies.
Specifically, these elongated blue structures may correspond to the blue,
collimated and extended emission--line regions (EELR's) commonly
observed in Seyferts (\eg\ Haniff \etal\ 1988, Wilson \etal\ 1988). The
suggestion of a strong causal relationship between nuclear activity and gas
motions within the innermost regions of a barred potential (Paper I and
references therein) would imply a correspondence between the orientations
of the EELR and any nuclear bar.

To investigate this possibility further, we have measured the alignments
of the (7) extended blue regions in our colour maps. They are compared
to the orientations of the nuclear isophotes in Table 5. There appears to be
a particularly close correspondence between these respective P.A.'s in NGC3412,
3945,
4536 and 4643. Even in the remaining cases, it is clear that there
exists a considerable misalignment between the extended blue nuclear regions
and the orientation of the primary bars. More extensive observational data
are required to further investigate this possible correspondence.

Near IR data for a small sample of active galaxies are presented by Hunt \&\
Giovanardi (1992). More detailed studies of the 2D near IR
luminosity distributions in LINERS/starbursts (Forbes \etal\ 1992) suggest
the nuclei of such galaxies to be redder than typical old stellar populations
-- particularly in (H--K). The colours of the LINERS/starbursts in Forbes
\etal\
are also consistent with those of Seyfert I nuclei (Kotilainen \etal\ 1992b,
Kotilainen \&\
Ward 1994), although the emission mechanisms are likely
to be quite different.
These red colours have most recently
been interpreted as implying re--radiation
from hot dust at 2.2$\mu$m, influenced by
the proximity to the active nucleus (Kotilainen \&\ Ward 1994). The
contribution
from M giants/supergiants is, however, likely to be an important factor in
starbursts.

Unfortunately, there appear considerable uncertainties regarding the colours
derived for these LINER/starburst and Seyfert I samples.
Estimates of the sky background are unreliable in data of such limited
spatial coverage. Furthermore,
the removal of a nuclear (non--stellar) contribution
to each passband is highly uncertain given the assumptions of azimuthal
symmetry imposed, difficulties in the assignment of a suitable PSF and the
unreliable nature of complex luminosity profile deconvolution given the
limited data available (Kotilainen \etal\ 1992a). The
apertures used by Kotilainen \&\ Ward (1994) are not sufficiently
small to eliminate significant contributions from circumnuclear star
formation in some objects. Interestingly, of the 4 Seyfert galaxies in
our sample (Table 1), only NGC4593
has an unusually red (H--K) colour. Our values of (J--H)=0.67
mag. and (H--K)=0.44 mag. for this object are in reasonable agreement with
those
of Kotilainen \&\ Ward (1994) over similar apertures.

Clearly, detailed studies of the near IR luminosity distributions of
LINER/starburst and Seyfert galaxies must await observations with high
spatial resolution
and large areal coverage.

\section{Conclusions}

In this paper, we present near IR photometry of 32 large, non--interacting
barred galaxies. Fully 21 of the 29 non edge--on objects in this sample
display measurable inner isophote twists with respect to the primary bars,
thereby confirming the frequency of this phenomenon inferred by Shaw \etal\
(1993). Such twists are identified in galaxies as late as SBbc.

Comparisons have been undertaken between the scale of such isophote twists and
the typical dimensions of nuclear bars, circumnuclear rings (CNR's)
and the inferred
maximum dimensions of the x2 orbital family within each primary bar. We
conclude
that 7 of these galaxies may possess a
secondary (nuclear) bar misaligned from the primary.
NGC4274, 4321 and 4643 present
the clearest evidence of this morphology. There appears to be a correlation
between the axis ratios of the inner and primary bars, as the nuclear bars are
much
thinner and more distinct in later type galaxies.
However, we find no evidence of a
correlation between twisted near IR isophotes and the presence of CNR's.
Twists in the
remaining galaxies may reflect the presence of triaxial bulges or of
oblate bulges misaligned with the principal bar.

The near IR colour distributions in our sample yield evidence for very blue
structures within the
nuclei of 19 galaxies -- these regions being
bluer than typical old stellar populations by
$\sim$0.30 mag. in (J--H) and $\sim$0.23 mag. in (H--K). These blue colours
do not appear to correlate with the presence of CNR's.

It is unlikely that these blue colours
result from a contribution from young giants/supergiants within unresolved
circumnuclear star forming rings as the CNR's observed in the present data
are all red and probably dominated by emission from M stars.
If such blue regions reflect
reddening variations within the nuclei for objects with typical
elliptical/bulge
colours, a {\it reduction} in extinction of A$_{v}$ $\sim$1.0 mag. is implied.
This may result from the geometry of the gas distribution within the corotation
radius
of the barred potential.

Interestingly, 3 of the 4 galaxies
showing clear evidence for resolved CNR's in our data also possess blue
nuclei -- one of these being a Seyfert, one a LINER and one a starburst. This
may be suggestive of a link with nuclear activity, although
the extended
nature of the blue nuclear colours, even in galaxies not classified as active,
argues against their resulting from the featureless blue continuum observed
in many Seyferts/LINERs. However, a general equivalence between the
orientations of these blue regions and the innermost nuclear isophotes
may suggest that the former are a manifestation of the extended, blue,
emission--line regions commonly identified in some Seyfert galaxies.

\section*{Acknowledgements}

We thank Dr. F. Combes and D. Friedli for valuable input. MAS was supported
by PPARC throughout the course of this work. We acknowledge support, in
the form of data analysis software, provided by the Starlink Project.

\newpage
\parindent 0.0cm
\section*{References}

Arsenault R., 1989. A \&\ A, 217, 66

Arsenault R., Boulesteix J., Georgelin Y., Roy J.--R., 1988. A \&\ A,
200, 29

Athanassoula E., 1992. MNRAS, 259, 328

Athanassoula E., Martinet L., 1980. A \&\ A, 87, L10

Baumgart C., Peterson B., 1986. PASP, 98, 56

Benedict G., Higdon J., Tollestrup E., Hahn J., Harvey P., 1992. AJ,
103, 757

Boer B., Schultz H., 1993. A \&\ A, 277, 397

Bottinelli L., Gouguenheim L., Paturel G., \&\ de Vaucouleurs G.,
1984. A \&\ A, 56, 381

Buta R., Crocker D., 1993. AJ, 105, 1344

Combes F., Gerin M., 1985. A \&\ A, 150, 327

Combes F., Gerin M., Nakai N., Kawabe R., Shaw M., 1992. A \&\ A, 259, L27

Devereux N., Kenney J., Young J., 1992. AJ, 103, 784

Elias J., \etal, 1982. AJ, 87, 1029

Ellis T., \etal, 1992. Proc. SPIE, 1765, 94

Forbes D., Ward M., DePoy D., Boisson C., Smith M., 1992. MNRAS, 254,
509

Friedli D., Martinet L., 1993. A \&\ A, 277, 27

Garcia--Barreto J., Downes D., Combes F., Gerin M., Magri C., Carrasco L.,
Cruz--Gonzalez I., 1991. A \&\ A, 244, 257

Gerin M., Nakai N., Combes F., 1988. A \&\ A, 203, 44

Griersmith D., Hyland A., Jones T., 1982. AJ, 87, 1106

Haniff C., Wilson A., Ward M., 1988. ApJ, 334, 104

Hawarden T., Mountain C., Leggett S., Puxley P., 1986. MNRAS, 221, 41p

Heckman T., 1980. A \&\ A, 88, 365

Hummel E., J\"{o}rs\"{a}ter S., Lindblad P., Sandqvist A., 1987a.
A \&\ A, 172, 51

Hummel E., van der Hulst J., Keel W., 1987b. A \&\ A, 172, 32

Hunt L. K., Giovanardi C., 1992. AJ, 104, 1018

Jarvis B., Dubath P., Martinet L., Bacon R., 1988. A \&\ A supp.,
74, 513

Jedrzejewski R., 1987. MNRAS, 226, 747

Keel W., 1983. ApJ suppl., 52, 229

Kenney J., Wilson C., Scoville N., Devereux N., Young J., 1992. ApJ,
395, L79

Kormendy J., 1979. ApJ, 227, 714

Kormendy J., 1982. ApJ, 257, 75

Kotilainen J., Ward M., Boisson C., DePoy D., Bryant L., Smith
M., 1992a. MNRAS, 256, 125

Kotilainen J., Ward M., Boisson C., DePoy D., Smith
M., 1992b. MNRAS, 256, 149

Kotilainen J., Ward M., 1994. MNRAS, 266, 953

Palumbo G., Tanzella--Nitti G., Vettolani G., 1980. Catalogue of
Radial Velocities, Gordon \&\ Breach, London

Pfenniger D., Norman C., 1990. ApJ, 363, 391

Pogge R., 1989. ApJ supp., 71, 433

Schommmer R., Caldwell N., Wilson A., Baldwin J., Phillips M,
Williams T., Turtle A., 1988. ApJ, 324, 154

Shaw M., Combes F., Axon D., Wright G., 1993. A \&\ A, 273, 31 (Paper I)

Shlosman I., Frank J., Begelman M., 1989. Nature, 338, 45

Sparke L., Sellwood J., 1987. MNRAS, 225, 653

Terlevich R., Melnick J., 1985. MNRAS, 213, 841

de Vaucouleurs G., de Vaucouleurs A., Corwin H., 1976. Second
Reference Catalogue of Bright Galaxies, University of Texas press

de Vaucouleurs G., Buta R., 1980. AJ, 85, 637

Veron--Citty M., Veron P., 1989. Catalogue of Active Galaxies and
Quasars, ESO scientific report no. 13

Wilson A., Ward M., Haniff C., 1988. ApJ, 334, 121

Wilson A., Braatz J., Heckman T., Krolik J., Miley G., 1993. ApJ,
419, L61

Winge C., Pastoriza M., Storchi--Bergmann T., Lipari S., 1992. ApJ, 393,
98

Yee H., 1983. ApJ, 272, 473

\newpage
\parindent 1.0cm
\normalsize
\section*{Figure captions}

\begin{description}
\item [Fig. 1 :] Contour plots from subregions of the J--band images of
NGC3941, 3953, 4643 and 4754. In all cases, N is down, E to the left. The
axes are marked in pixels (1 pixel = 1.4 arcsec) and the data are
contoured from 14.0 to 18.4 mag. arcsec$^{-2}$ in 0.4 mag. arcsec$^{-2}$
intervals. The primary bar position angles, derived as discussed in the text,
are shown by the dashed lines.

\item [Fig. 2 :] Plots of the radial variation of position angle,
ellipticity, $\cos 4\theta$ and azimuthally averaged surface brightness
($\mu$)
derived from ellipse fitting to the galaxies displayed in Figure 1. The
profiles
for each passband are distinguished by the symbols shown in the legend
accompanying the lower right hand panel for each galaxy. The horizontal
dashed lines in the position angle profiles denote the P.A. of the primary
bars,
whilst the arrow identifies the inferred minor axis dimensions of these bars.

\item [Fig. 3 :] As Figure 1, but for NGC4274 (a) and NGC4321 (b). An expanded
plot of the central region of (b) is given in (c). Data in (a)
are contoured using the levels adopted in Figure 1. In (b)
contours are plotted from 14.0
to 17.6 (J) mag. arcsec$^{-2}$ in 0.4 mag. arcsec$^{-2}$ intervals, whilst in
(c)
contours are from 14.0 to 16.2 mag. arcsec$^{-2}$ in 0.2 mag. arcsec$^{-2}$
intervals.

\item [Fig. 4 :] As Figure 2, but for NGC4321.

\item [Fig. 5 :] Greyscale plots of the 2D colour distributions within:
(a) NGC3351 (J--K); (b) NGC3945 (H--K); (c) NGC4303 (J--H); (d) NGC4321 (J--H)
and (e) NGC4754 (J--H).
Display levels are: (a) 0.5 (black) to 1.0 (white) mag.; (b) 0.0 to 0.35 mag.;
(c) 0.35 to 0.7 mag.; (d) 0.25 to 1.2 mag. and (e) 0.3 to 0.65 mag. The
spatial coverage is: (a) 91x92 arcsec; (b) 117x109 arcsec; (c) 127x139 arcsec;
(d) 206x186 arcsec and (e) 104x135 arcsec.

\item [Fig. 6 :] Near IR colour-colour diagram of the present sample.
Also displayed is the region occupied by typical old stellar populations
(circle),
and the track delineated by a $\sim$30 \%\ contribution from a
600 K black body. The influence of a 10\%\ contribution from hot (T=200K)
dust at 2.2$\mu$m and a reddening vector of A$_v$ = 1 mag. are also shown. The
galaxies in the present sample are plotted according to their classifications
inferred from their emission--line spectra in Keel (1983). Typical
uncertainties
on both colours are plotted in the lower right corner. Galaxies identified as
Seyferts in Veron--Cetty \&\ Veron (1989) are marked by filled symbols.

\end{description}

\section*{Table captions}

\begin{description}
\item [Table 1 :] Properties of the present barred galaxy sample.

\item [Table 2 :] Summary of results derived from ellipse fitting to
the sample galaxies.

\item [Table 3 :] Results derived from analyses of the colour
distributions in the sample galaxies.

\item [Table 4 :] Colour indices for the nuclear rings in NGC3351, 4303,
4314 and 4321. Ring dimensions are quoted as major x minor axes, derived
from the (J--H) colour maps or, in the case of NGC3351, (J--K).

\item [Table 5 :] Comparison of the orientations of the extended blue regions
in the colour maps and the nuclear isophote orientations listed in Table 2.

\end{description}

\newpage
\section*{Notes to Table 1}

\begin{description}
\item [source of type :] RC2 -- de Vaucouleurs, de Vaucouleurs \&\ Corwin
(1976);
dVB -- de Vaucouleurs \&\ Buta (1980).

\item [dist. :] Object distance, taken from recessional velocities in
Palumbo, Tanzella--Nitti \&\ Vettolani (1980), using the distance moduli in
Bottinelli {\it et al.} (1984), or -- for NGC4314 -- the distance estimated
by Garcia--Barreto et al. (1991). An
H$_{\circ}$ of 50 km s$^{-1}$ Mpc$^{-1}$ is assumed throughout.

\item [classification :] Object classification from inclusion within the
starburst/A.G.N. or control samples of Arsenault (1989).

\item [comments :] nuc. ring -- nuclear ring; opt. twists -- twists in optical
isophotes with radius.

\item [source of comments :] BC93 -- Buta \&\ Crocker (1993); VV89 --
Veron--Citty \&\ Veron (1989); D92 -- Devereux, Kenney \&\ Young (1992);
K79 -- Kormendy (1979); J88 -- Jarvis {\it et al.} (1988); A88 -- Arsenault
{\it et al.} (1988).

\end{description}

\newpage
\section*{Notes to Table 2}

\begin{description}
\item [col. 2 :] A = anticlockwise, C = clockwise, measured from the POSS
or the present dataset assuming spiral arms are trailing.

\item [cols. 3--6 :] POSS corresponds to position angles defined
from the POSS survey plates, whilst ``ellipse fit'' lists those derived
from ellipse fitting to the present dataset (averages of all 3 near IR
frames with the exception of NGC3945
which is measured from the J--band image alone).

\item [col. 7 : ] Galaxy inclination angles derived from log R$_{25}$
measures in de Vaucouleurs, de Vaucouleurs \&\ Corwin (1976) assuming
an intrinsic disc axial ratio of 0.2.

\item [cols. 8, 9 :] Offset between inner near--IR isophotes and those
of the principal bar component. Measures derived from ellipse fitting
results except for NGC3945, NGC4340 and NGC4371 where $\theta$ is measured
directly from the images. For NGC3351, values refer to
the J and H images only.
Sign of $\theta$ identifies whether the inner
misaligned isophotes lead (+) or trail (--) the primary bar. In galaxies
for which the sense of rotation is undefined, the isophote twists are
denoted by ``$\pm$''.

\item [cols. 10, 11 :] Spatial scale over which the isophote twists are
observed.

\item [col. 12 :] Radius of the primary bar,
measured from the J--band images except for
NGC4321 which comes from the estimate of Arsenault et al. (1988).

\item [col. 13 :] The spatial scale of the misaligned isophotes ($\delta$R)
expressed as a fraction of the radius (L) of the primary bar.

\end{description}

\newpage
\section*{Notes to Table 3}

\begin{description}
\item [col. 2 :] Aperture radius (kpc) over which mean colour
indices in cols. 3,4 correspond. For the ``control'' sample, an
aperture of 2 pixels was adopted throughout.

\item [cols. 3, 4 :] Colour indices with errors corresponding to standard
deviations about these mean values within the specified apertures.

\item [cols. 5, 6 :] Magnitudes of colour differences between
nuclear structures in colour maps and the colours of the surrounding
regions.

\item [col. 7 :] Comments on structures evident in colour maps.

\end{description}

\newpage
\section*{Notes to Table 5}

\begin{description}
\item [col. 2 :] Colour map used in evaluation of entries in cols. 3 and 4.
All corresponds to J--H, H--K and J--K colour maps.

\item [col. 3 :] Position angle of the elongated blue structures evident
in the colour maps cited in col. 2. Where several colour maps have been
used, values quoted are weighted means.

\item [col. 4 :] Differences between PA$_{col}$ in col. 3
and the position angle of the primary bar (as defined in Table 2). Measured
in the sense PA$_{col}$ -- PA (primary bar).

\item [col. 5 :] $\theta$ measures in Table 2, expressed in the sense
PA (inner isophotes) -- PA (primary bar). The sign of $\theta$ may differ
from that in Table 2 as the latter measures identify whether the inner
isophotes trail or lead the primary bars.

\end{description}
\end{document}